\documentclass[a4paper,11pt]{article}
\usepackage{pos}

\title{Adjoint chromoelectric correlators for heavy quarkonium diffusion}

\author*[a,b]{Julian Mayer-Steudte}

\affiliation[a]{Technical University of Munich, TUM School of Natural Sciences, Physics Department, James-Franck-Strasse 1, 85748 Garching, Germany}
\affiliation[b]{Munich Data Science Institute, Technische Universit\"at M\"unchen, \\
Walther-von-Dyck-Strasse 10, 85748 Garching, Germany}

\emailAdd{julian.mayer-steudte@tum.de}

\abstract{We here measure, for the first time, adjoint chromoelectric correlators at finite temperatures that encode the diffusion of quarkonium in the medium. Understanding the dynamics of quarkonium in the QGP plays an essential role in understanding quarkonium suppression and the QGP in general. We perform SU(3) gauge theory calculations and use gradient flow to improve the signal-to-noise ratio and chromoelectric field discretizations. The continuum limit and the zero-flow-time extrapolation are performed, and the final result is compared with perturbative results. We observe that the correlators at a high temperature are well described by the perturbative form; furthermore, we observe multiplicative scaling of the adjoint correlators with respect to the fundamental correlator describing heavy quark diffusion.}

\FullConference{The 41st International Symposium on Lattice Field Theory (LATTICE2024)\\
 28 July - 3 August 2024\\
Liverpool, UK\\}

\allowdisplaybreaks
\begin{document}
\maketitle

\section{Introduction}

Quantum chromodynamics (QCD) is the theory that describes the strong force mediated through gluons between three families of quarks with increasing mass. Confinement is a distinguished property of QCD, which describes bound states like protons and nuclei, as well as the bound state of two heavy quarks known as quarkonium. However, there is a phase transition to a deconfinement phase called quark-gluon plasma (QGP) at higher temperatures. Understanding the QGP is crucial to achieving a better understanding of the early universe and the dynamics of heavy ion collisions. Studying quarkonium as a probe of the QGP has been proposed theoretically in~\cite{Matsui:1986dk}, and experimental programs have emerged - see Refs.~\cite{Aarts:2016hap,Zhao:2020jqu,Brambilla:2014jmp}.

A direct approach to QCD to describe quarkonium dynamics is unfeasible, and developing an effective field (EFT) approach is more appropriate since it has a distinct scale hierarchy: the heavy quark mass $M$, the relative momentum $Mv$, and the binding energy, with $v$ being the heavy quark velocity inside the quarkonium~\cite{Brambilla:1999xf,Brambilla:2004jw}. Integrating out the scale $M$ leads to the heavy quark effective theory (HQET), or non-relativistic QCD (NRQCD)~\cite{Caswell:1985ui,Thacker:1990bm}, respectively. Integrating out the momentum transfer $Mv$ gives potential NRQCD (pNRQCD). A pNRQCD description can be extended to finite temperature cases using an open quantum system approach based on Lindblad equations~\cite{Brambilla:2016wgg,Brambilla:2017zei}.

Refs.~\cite{Moore:2004tg,Caron-Huot:2008dyw,Bouttefeux:2020ycy} propose studying the diffusion of a heavy quark. The momentum diffusion coefficient $\kappa$ is encoded in the infrared limit of the spectral function of a chromoelectric correlator in the leading order in the heavy quark mass expansion. The first-order correction contribution is connected to a chromomagnetic correlator. The leading contribution to $\kappa$ is determined in quenched theory in~\cite{Banerjee:2011ra,Francis:2015daa,Brambilla:2020siz}, in~\cite{Altenkort:2020fgs} the gradient flow method~\cite{Narayanan:2006rf,Luscher:2009eq,Luscher:2010iy} was for the first time employed to improve the chromoelectric field non-perturbatively. In our previous study~\cite{Brambilla:2022xbd}, we computed the next order correction from the chromomagnetic correlator utilizing the gradient flow method. The first steps towards computations with 2+1 dynamical fermions were performed in~\cite{Altenkort:2023oms}.

As obtained from a pNRQCD description, the quarkonium diffusion coefficients, which enter the Lindblad equations, are encoded in chromoelectric correlators connected by adjoint Wilson lines. Although these correlators and transport coefficients are of high importance for the study of quarkonium in QGP and as input for the quarkonium production in heavy ion collisions~\cite{Brambilla:2016wgg,Brambilla:2017zei,Yao:2020eqy,Scheihing-Hitschfeld:2023tuz,Binder:2021otw,Brambilla:2024tqg,Brambilla:2022ynh}, they have not yet been calculated non-perturbatively on the lattice at finite temperatures. We therefore aim to compute and present the results of adjoint chromoelectric correlators on the lattice by utilizing the gradient flow algorithm in quenched theory. This work was recently published in~\cite{Brambilla:2025cqy}.

\section{Adjoint chromoelectric correlators}

We construct three distinguishable correlators, where chromoelectric fields are attached to a Wilson line in the adjoint representation:
\begin{align}
    G_{E}(\tau) &= -\frac{1}{3}\sum_{i=1}^3 \langle E_{i,a}(\tau) U^{\mathrm{adj}}_{ab}(\tau,0) E_{i,b}(0)\rangle, \label{NonSymCorr}\\
    G_{E}^{\mathrm{oct}}(\tau) &= -\frac{1}{3\langle l_8\rangle}\sum_{i=1}^3 \langle U^{\mathrm{adj}}_{ea}(1/T,\tau)d_{abc}E_{i,c}(\tau) U^{\mathrm{adj}}_{bd}(\tau,0)d_{def}E_{i,f}(0)\rangle,\label{octCorr}\\
    G_{E}^{\mathrm{sym}}(\tau) &= \frac{1}{3\langle l_8\rangle} \sum_{i=1}^3 \langle U^{\mathrm{adj}}_{ea}(1/T,\tau)f_{abc}E_{i,c}(\tau) U^{\mathrm{adj}}_{bd}(\tau,0)f_{def}E_{i,f}(0)\rangle,\label{symCorr}
\end{align}
where $G_{E}^{\mathrm{sym}}$ can be seen as the adjoint version of the fundamental chromoelectric correlator
\begin{equation}
    G_{E}^{\mathrm{fund}}(\tau) = -\sum_{i=1}^3 \frac{\langle \mathrm{Re}\mathrm{Tr} [U(1/T,\tau)E_{i}(\tau)U(\tau,0)E_{i}(0)]\rangle}{3\langle l_3\rangle}, \label{fundCorr}
\end{equation}
which describes the diffusion of a heavy quark. $U^\mathrm{adj}_{ab}(\tau_1,\tau_2)$ is the temporal adjoint Wilson line with adjoint color indices $a,b$ connecting timeslice $\tau_1$ to $\tau_2$ and is defined through the fundamental temporal Wilson line $U(\tau_1,\tau_2)$ as
\begin{align}
    U^{\mathrm{adj}}_{ab}(\tau_1,\tau_2) &= \frac{1}{2}\mathrm{Tr}[U(\tau_1,\tau_2)\lambda_aU(\tau_2,\tau_1)\lambda_b]\label{eq:adjoint_wilson_line_definition},
\end{align}
where $\lambda_a$ are the Gell-Mann matrices, $E_{i,a}(\tau)$ is the chromoelectric field with spatial component $i$ and color index $a$ and is defined through a field component in the fundamental representation $E_i$ as
\begin{align}
    E_{i,a} &= \mathrm{Tr}[\lambda_aE_i]\label{eq:adjoint_field_definition}.
\end{align}
$d_{abc}$ and $f_{abc}$ are, respectively, the symmetric and antisymmetric structure constants. $l_3$ and $l_8$ are the non-trace-normalized Polyakov loop in fundamental and adjoint representation:
\begin{align}
    l_3 &= \mathrm{Tr}U(1/T,0)\\
    l_8 &= \mathrm{Tr}U^\mathrm{adj}(1/T,0) = \sum_{a=1}^8U_{aa}^\mathrm{adj}(1/T,0) = |l_3|^2-1
\end{align}
with $T$ being the temperature. $\langle ...\rangle$ denotes the path integral expectation value, i.e., the average over the gauge field ensemble.

We employ the clover and a two-plaquette discretization of the field components, which we label CLO and 2PL, respectively. The clover discretization of the field strength tensor $F_{\mu\nu}$ is given by the sum of four plaquettes
\begin{align}
    a^2F_{\mu\nu} &= -\frac{i}{8}(Q_{\mu\nu}-Q_{\nu\mu}), \\
    Q_{\mu\nu} &= U_{\mu,\nu} + U_{\nu,-\mu} + U_{-\mu,-\nu} + U_{-\nu,\mu} = Q_{\nu\mu}^\dagger, 
\end{align}
with $U_{\mu,\nu}$ being the plaquette in the $\mu-\nu$-plane and $a$ the lattice spacing. The chromoelectric fields are then given by $E_i = -F_{i,4}$. The 2PL discretization includes only two of the four plaquettes such that the chromoelectric field components have no overlap of temporal link variables in the $G_E(\tau)$ correlator. For $G_E^\mathrm{oct}(\tau)$ and $G_E^\mathrm{sym}(\tau)$, the explicit choice of the two plaquettes is symmetric as long as the start and end of the correlator are set at the half of the temporal extend of the operator.

The finite-extension discretization of the chromoelectric field causes gluonic self-interactions, which give logarithmic behaviors to the continuum limit and, hence, impedes the conducting of reliable continuum limits. Therefore, we rely on the gradient flow method~\cite{Narayanan:2006rf,Luscher:2009eq,Luscher:2010iy} to improve the chromoelectric field insertions. In addition, it improves our signal-to-noise ratio and allows us to reach results at larger separations $\tau$. However, gradient flow introduces an additional fictitious dimension called flow time $\tau_F$ in units of $a^2$ to any observable, which eventually has to be removed  by a zero-flow-time extrapolation. In addition, the flow time acts as a regulator for any divergence.

We understand that $G_E(\tau)$ is related to the singlet-octet transition, describing the transition from a bound state to a scattering state (dissociation and recombination), and $G_E^\mathrm{oct}(\tau)$ to the octet-octet transition, describing the transition from one scattering state to another. In addition, $G_E^\mathrm{sym}(\tau)$ describes the diffusion of a heavy adjoint source.

An operator with a Wilson line of length $\tau$ comes with a mass divergence $\delta m\propto 1/\sqrt{8\tau_F}$ where we explicitly observe the flow radius as the regulator of the divergence. The operators $G_E^\mathrm{oct}(\tau)$ and $G_E^\mathrm{sym}(\tau)$ are Polyakov loops with chromoelectric field insertions separated by $\tau$; hence, the divergence is given by $e^{-\delta m/T}$ which is canceled by the Polyakov loop normalization, which carries the same divergence. In contrast, $G_E(\tau)$ has an explicit divergence as $e^{-\delta m \tau}$. We renormalize this operator by employing the renormalization condition from~\cite{Gupta:2007ax} with the renormalized and trace-normalized Polyakov loop $L_8^r$ as
\begin{align}
    L_8^r = \frac{1}{8}e^{\delta m(\tau_F)/T}\langle l_8(\tau_F)\rangle
\end{align}
where we include the flow-time dependence explicitly. We solve for the divergence and define the renormalized observable as
\begin{align}
    G_E^r(\tau, \tau_F) &= e^{\delta m(\tau_F)\tau}G_E(\tau,\tau_F)\\
    &= \left( \frac{8 L_8^r}{\langle l_8(\tau_F)\rangle}\right)^{\tau T} G_E(\tau,\tau_F)
\end{align}
where $G_E(\tau,\tau_F)$ is the bare measurement on the lattice at finite flow time.

The three correlators $G_E^r$, $G_E^\mathrm{oct}$, and $G_E^\mathrm{sym}$ are divergence-free and hence we can perform reliable continuum and zero-flow-time limits. In preparation, we apply a tree-level improvement as
\begin{align}
    G_E^{\mathrm{imp},p} = \frac{G_E^p(0,\tau T)|_\mathrm{LO}}{G_E^p(0,\tau T)|_{LO}^\mathrm{lat}} G_E^{\mathrm{measured},p}(\tau_F,\tau T)
\end{align}
with $p=\{ r,\mathrm{oct},\mathrm{sym}\}$, $G_E^p(0,\tau T)|_\mathrm{LO}$ the tree-level result of the correlators in continuum, and $G_E^p(0,\tau T)|_{LO}^\mathrm{lat}$ the tree-level result in lattice perturbation theory. The equations are given by
\begin{align}
    \frac{G^{\mathrm{fund}}_E(\tau)\vert_{\mathrm{LO}}}{g^2C_F} &\equiv f(\tau) = \pi^2T^4\left[\frac{\cos^2(\pi\tau T)}{\sin^4(\pi\tau T)} + \frac{1}{3\sin^2(\pi\tau T)}\right],\label{eq:cont_normalization}\\
    \frac{G_E(\tau)\vert_{\mathrm{LO}}}{g^2 C_F} &= 2Nf(\tau),\label{eq:G_E_LO}\\
    \frac{G^{\mathrm{oct}}_E(\tau)\vert_{\mathrm{LO}}}{g^2 C_F} &= 2\frac{N^2-4}{N^2-1}f(\tau),\label{eq:G_E_oct_LO}\\
    \frac{G^{\mathrm{sym}}_E(\tau)\vert_{\mathrm{LO}}}{g^2 C_F} &= \frac{C_A}{C_F}f(\tau),\label{eq:G_E_sym_LO}\\
    f(\tau)\vert_{\mathrm{lat}} &= \frac{1}{3a^4} \int_{-\pi}^\pi \frac{d^3q}{(2\pi)^3}\frac{\cosh[\bar{q} N_\tau (\frac{1}{2}-\tau T)]}{\sinh (\bar{q} N_\tau/2)} \frac{1}{\sinh (\bar{q})} \begin{cases} \;
    \left( 1 + \frac{\tilde{q}^2}{4} \right) \, \left(\tilde{q}^2 - 
    \frac{(\tilde{q}^2)^2}{8} + \frac{\tilde{q}^4}{8} \right) & (\mathrm{CLO}) \\
    \left( \tilde{q}^2 + \frac{\tilde{q}^4 - \left( \tilde{q}^2 \right)^2}{8}
    \right) & (\mathrm{2PL})\end{cases}.
\end{align}
with $\bar{q} = 2\arcsin\left(\sqrt{\tilde{q}^2}/2\right)$, $\tilde{q}^n = \sum_{i=1}^3 2^n\sin^n\left( q_i/2\right)$, $N=3$, $C_F=((N^2 - 1) T_f) / N$, $T_f=1/2$, and $C_A=N$. The continuum result can be found in~\cite{Caron-Huot:2009ncn}.

\section{Lattice results}
\begin{table}
\caption{The simulation parameters for the gradient flow lattices.}
\label{tab:simulation_parameters}
    \centering
    \begin{tabular}{ccccc}
        $T/T_c$ & $N_\tau$ & $N_\mathrm{s}$ & \multicolumn{1}{c}{$\beta$} & $N_\mathrm{conf}$\\\hline
         1.5 & 16 & 48 & 6.872 & 1000 \\
             & 20 & 48 & 7.044 & 2002 \\
            & 24 & 48 & 7.192 & 2060 \\
            & 28 & 56 & 7.321 & 1882 \\
            & 34 & 68 & 7.483 & 1170
    \end{tabular}
    \hspace{1cm}
    \begin{tabular}{ccccc}
        $T/T_c$ & $N_\tau$ & $N_\mathrm{s}$ & \multicolumn{1}{c}{$\beta$} & $N_\mathrm{conf}$\\\hline
       10000 & 16 & 48 & 14.443 & 1000 \\
            & 20 & 48 & 14.635 & 1178 \\
            & 24 & 48 & 14.792 & 998 \\
            & 34 & 68 & 15.093 & 599
    \end{tabular}
\end{table}
We generate quenched SU(3) gauge field configurations at $T=1.5T_c$ and $T=10^4T_c$. We use Wilson action with overrelaxation and heat bath algorithm to sample the configurations; the simulation parameters are listed in Table~\ref{tab:simulation_parameters}, and the $\beta$-parameter was fine-tuned via the scale setting in~\cite{Francis:2015lha}. We use the publicly available \href{https://web.physics.utah.edu/~detar/milc/milcv7.html}{MILC} code. We pick the Symanzik action to solve the gradient flow numerically~\cite{Fritzsch:2013je, Bazavov:2021pik}, where we vary the step-sizes with fixed step-sizes among an ensemble.

\begin{figure}
    \centering
    \includegraphics[width=0.45\textwidth]{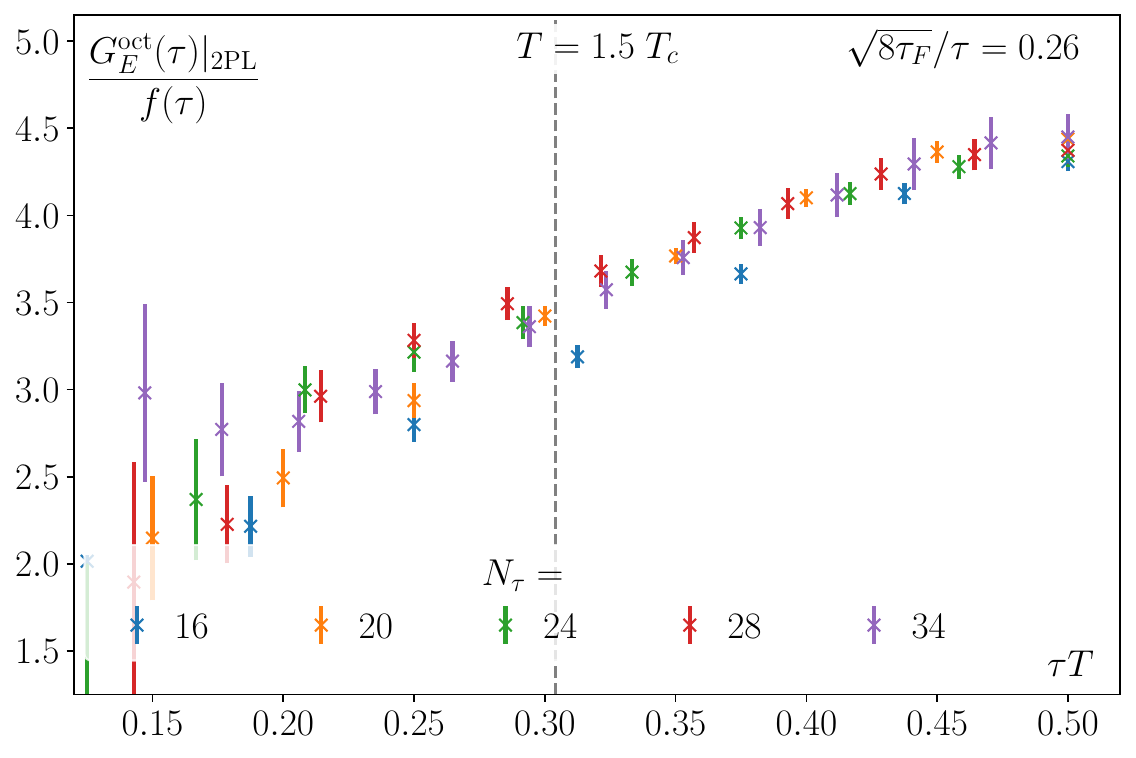}
    \includegraphics[width=0.45\textwidth]{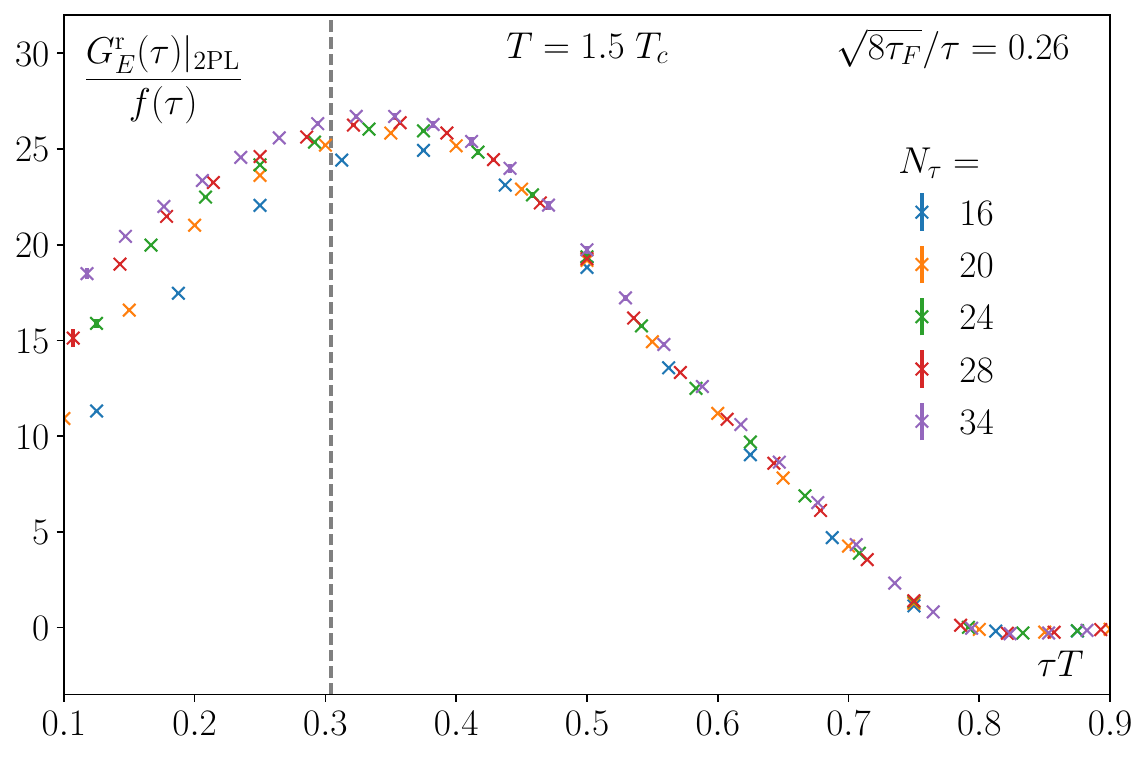}
    \caption{$G_E^r$ and $G_E^\mathrm{oct}$ with the 2PL operator, normalized with $f(\tau)$, at a fixed flow time ratio at $T=1.5T_c$ for all five lattice spacings.}
    \label{fig:G_E_rAoctet}
\end{figure}

\begin{figure}
    \centering
    \includegraphics[width=0.45\textwidth]{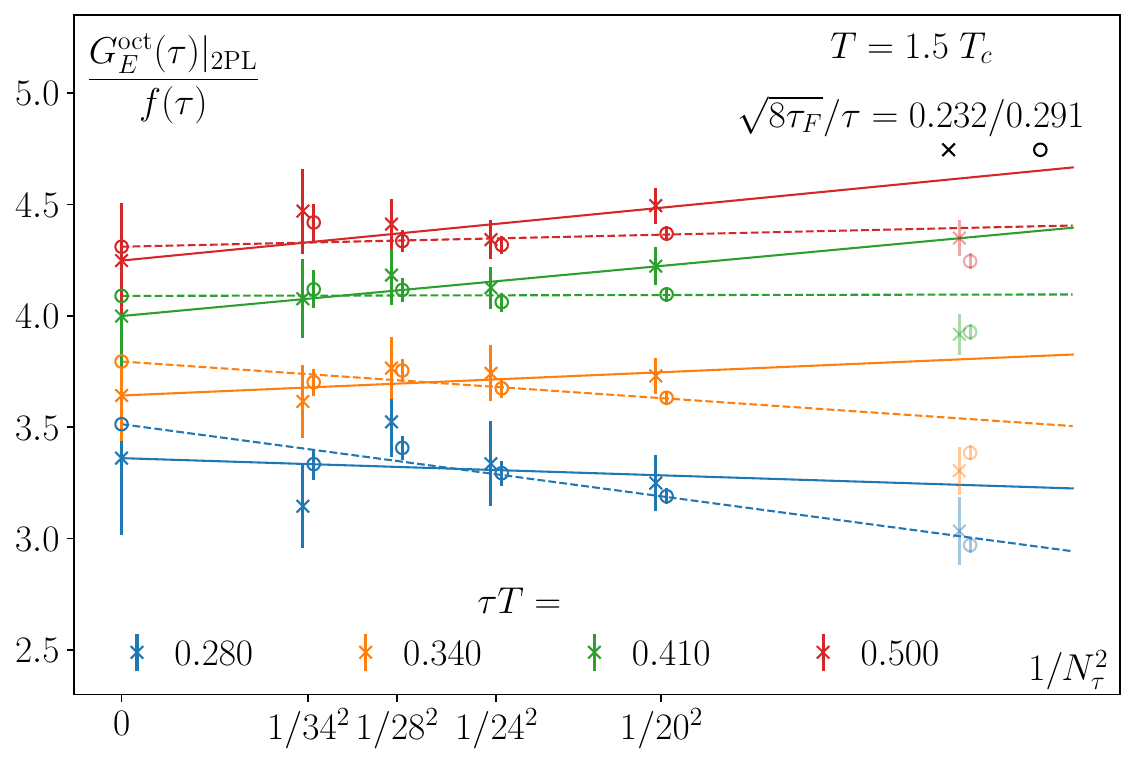}
    \includegraphics[width=0.45\textwidth]{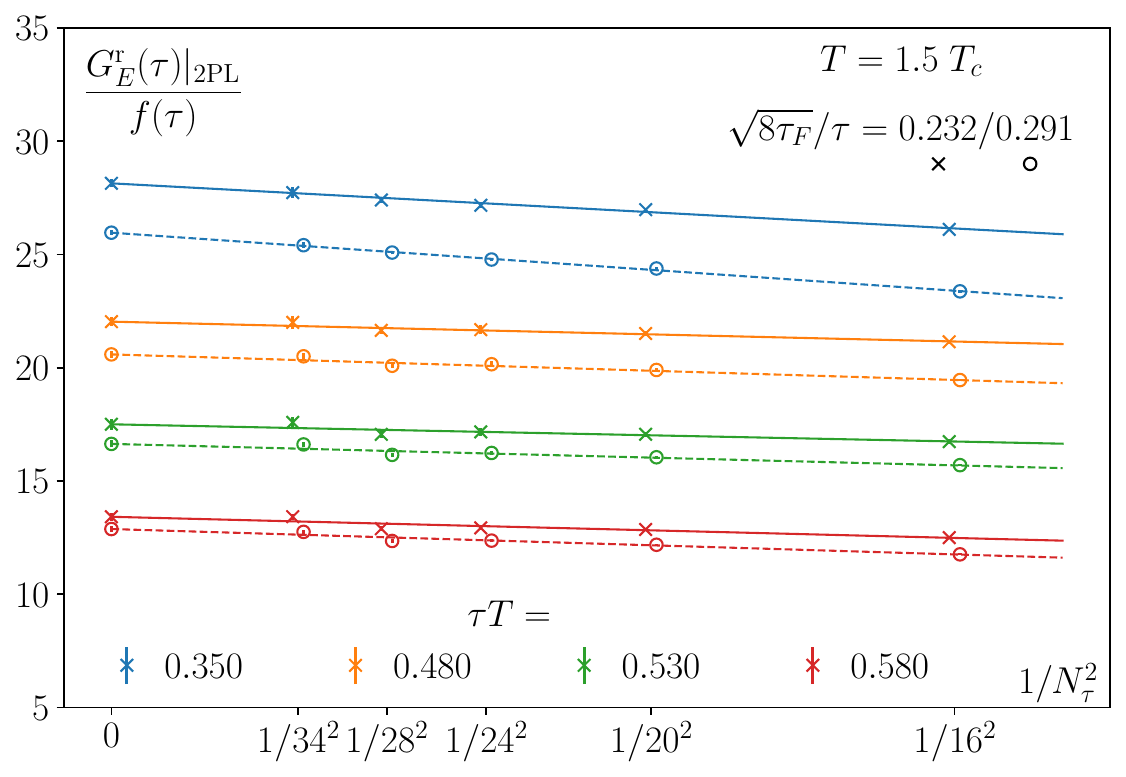}
    \caption{Examples of the continuum limit of $G_E^r$ and $G_E^\mathrm{oct}$ with the 2PL operator at $T=1.5T_c$. Dimmed data points are excluded from the continuum limit.}
    \label{fig:G_E_rAoctet_continuum_limit_example}
\end{figure}

Fig.~\ref{fig:G_E_rAoctet} shows $G_E^r$ and $G_E^\mathrm{oct}$ at a given flow time ratio at $T=1.5T_c$. $G_E^\mathrm{oct}$ and $G_E^\mathrm{sym}$ are symmetric around $\tau T=0.5$ by construction; conversely, $G_E$ is not. In preparation for the continuum limit, we interpolate the intermediate data points between the discrete $\tau T$ separations with cubic splines. Fig.~\ref{fig:G_E_rAoctet_continuum_limit_example} shows examples of the continuum limit for the $G_E^\mathrm{oct}$ and $G_E^r$ correlators. The continuum extrapolation is performed as a linear fit in $1/N_\tau^2\propto a^2$ since we use the Wilson action, which has a leading $\mathcal{O}(a^2)$ discretization artifact. For $G_E^\mathrm{oct}$, we exclude the $N_\tau=16$ ensembles from the continuum limit at $T=1.5T_c$ to keep $\chi^2/\mathrm{dof}$ close to 1, while for $T=10^4T_c$ all ensembles are included while delivering a good value for $\chi^2/\mathrm{dof}$.

\begin{figure}
    \centering
    \includegraphics[width=0.45\textwidth]{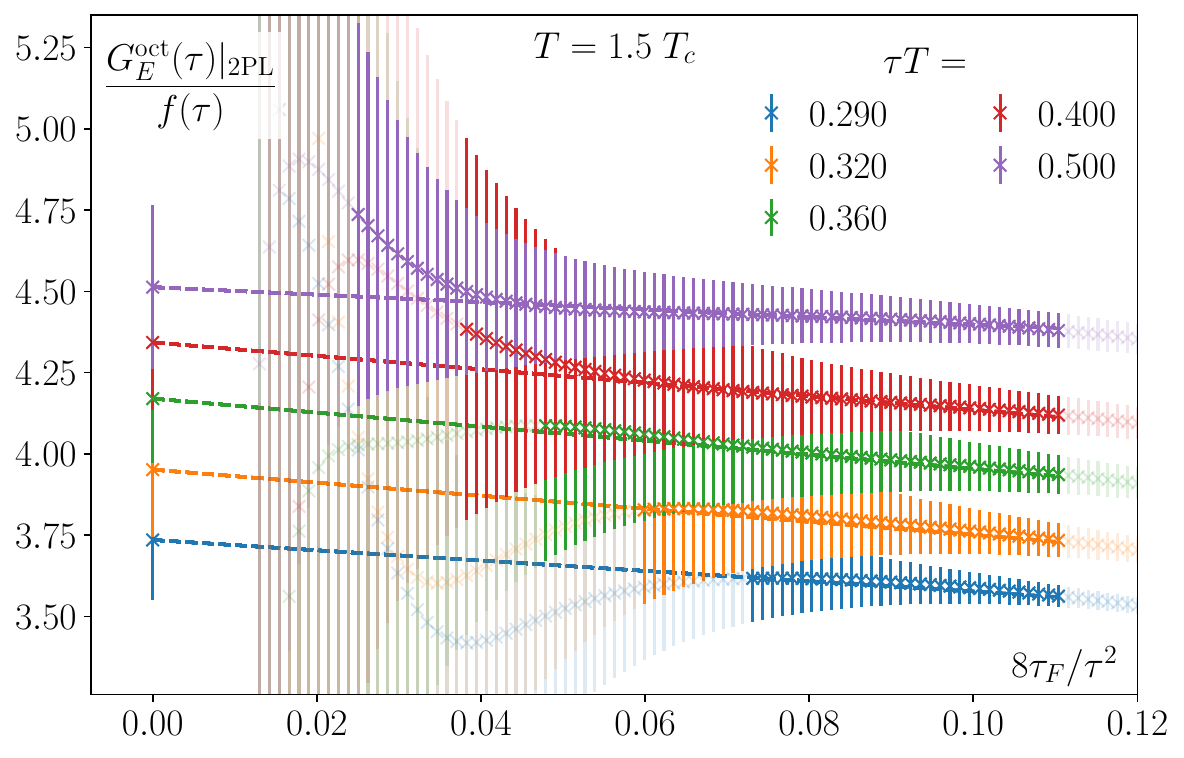}
    \includegraphics[width=0.45\textwidth]{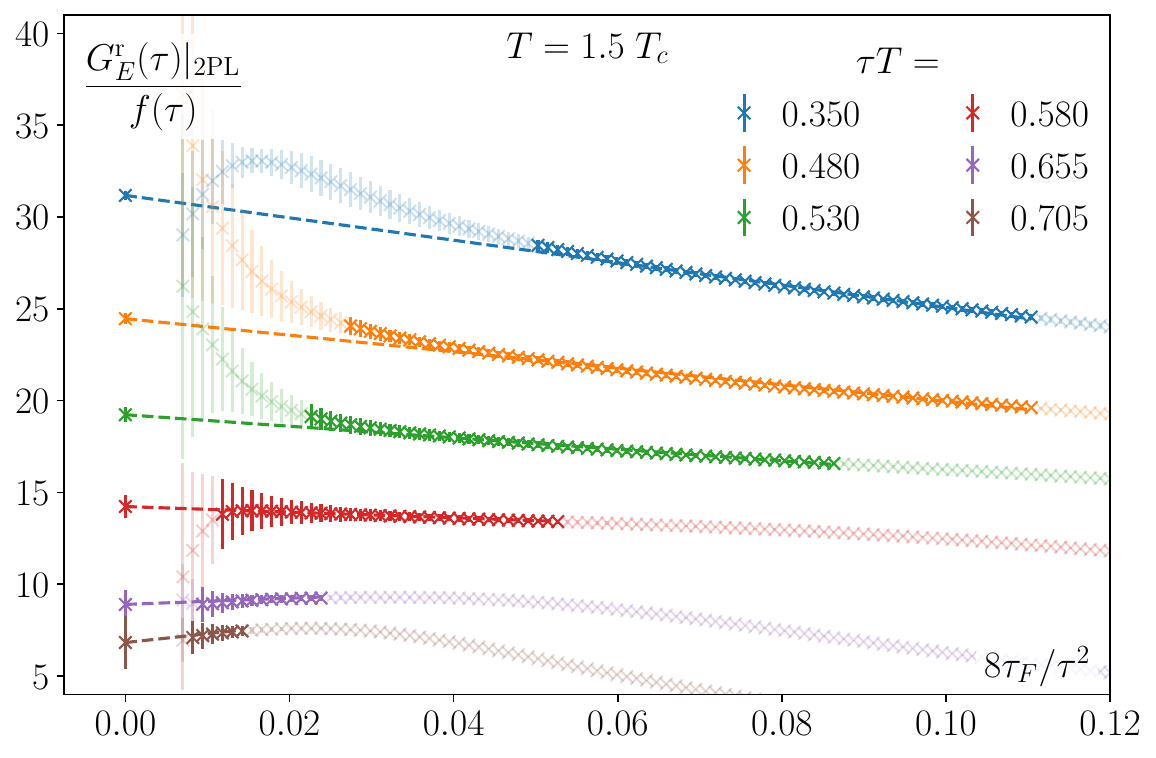}
    \caption{Examples of the zero-flow-time limit of $G_E^r$ and $G_E^\mathrm{oct}$ with the 2PL operator at $T=1.5T_c$. Dimmed data points are excluded from the zero-flow-time limit.}
    \label{fig:G_E_rAoctet_zftl_example}
\end{figure}

\begin{figure}
    \centering
    \includegraphics[width=0.45\textwidth]{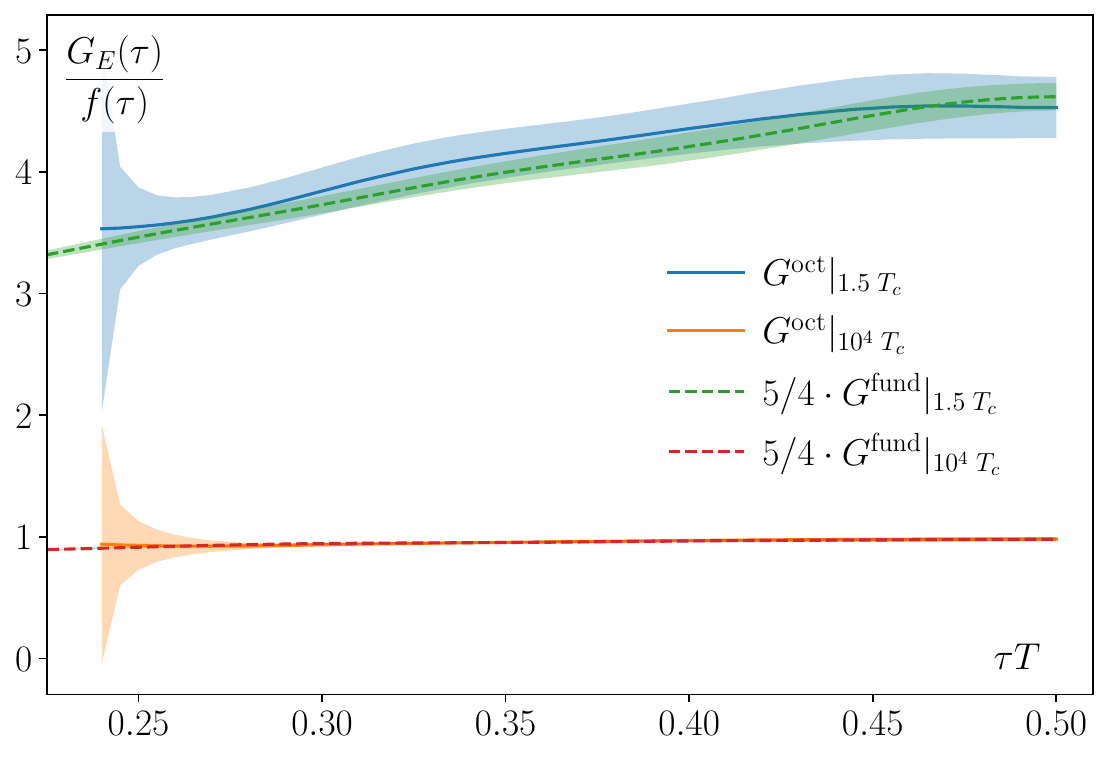}
    \includegraphics[width=0.45\textwidth]{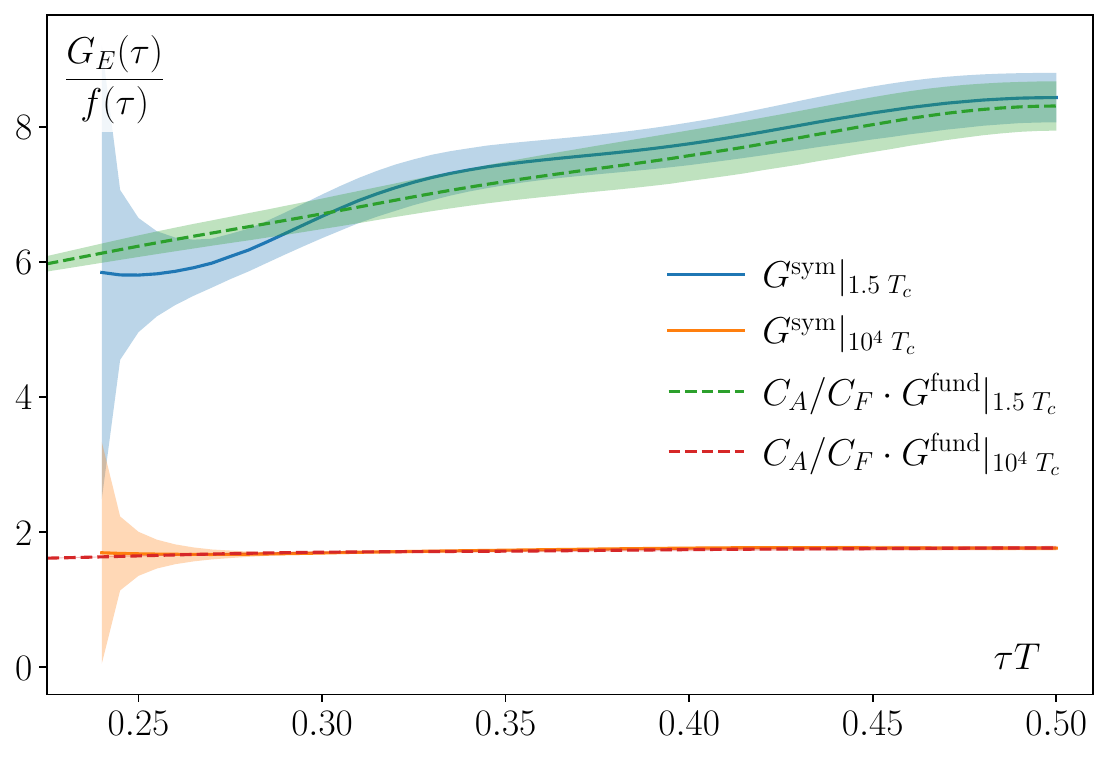}
    \caption{Final result of the symmetrized correlators (oct and symm) normalized with Eq.~\eqref{eq:cont_normalization}. We compare the result with the fundamental correlator scaled with the factor at LO given in Eqs.~\eqref{eq:G_E_LO},~\eqref{eq:G_E_oct_LO}, and~\eqref{eq:G_E_sym_LO}.}
    \label{fig:G_E_symmetric_final}
\end{figure}

In the next step, we perform the zero-flow-time limit by a linear in-flow-time extrapolation. We have shown in~\cite{Brambilla:2023fsi} that for flow times with $\sqrt{8\tau_F}>a$ the gradient flow method renormalizes the chromoelectric field effectively. From perturbative calculations at tree level~\cite{Altenkort:2020fgs} we know that the correlator is stable up to flow times $\sqrt{8\tau_F}<\tau/3$. Therefore, we restrict our considered flow time range to
\begin{align}
    a \leq \sqrt{8\tau_F} \leq \frac{\tau}{3}.
\end{align}
Examples of the zero-flow-time limit for $G_E^r(\tau)$ and $G_E^\mathrm{oct}(\tau)$ are presented in Fig.~\ref{fig:G_E_rAoctet_zftl_example}. We identify a linear flow-time behavior within the considered flow-time range and the error bands, where we perform the zero-flow-time extrapolation. Fig.~\ref{fig:G_E_symmetric_final} shows the final result of the symmetrized correlators (oct and sym) compared to the fundamental correlator. We observe that the adjoint correlators are scaled versions of the fundamental ones with the scaling factors given by LO results in Eqs.~\eqref{eq:G_E_oct_LO} and~\eqref{eq:G_E_sym_LO}. For $G_E^\mathrm{oct}$, the scaling factor is given by $5/4$, and for $G_E^\mathrm{sym}$ it is the Casimir scaling $C_A/C_F$. This observation simplifies an extraction of the heavy quark diffusion coefficient, which is given as the infrared limit of the spectral function of the correlators. Due to the linear scaling of the correlators, we expect $\kappa^\mathrm{oct}=(5/4)\kappa^\mathrm{fund}$ and $\kappa^\mathrm{sym}=(C_A/C_F)\kappa^\mathrm{fund}$ where $\kappa$ represents the diffusion coefficient.

\begin{figure}
    \centering
    \includegraphics[width=0.45\textwidth]{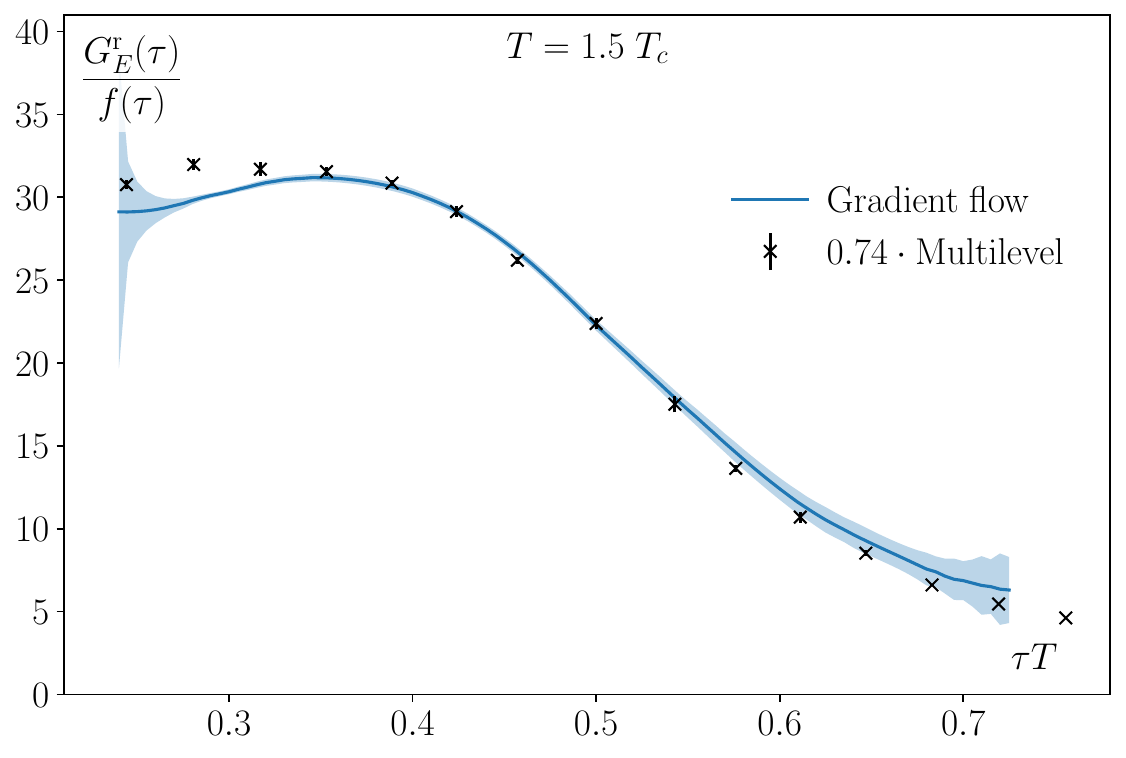}
    \includegraphics[width=0.45\textwidth]{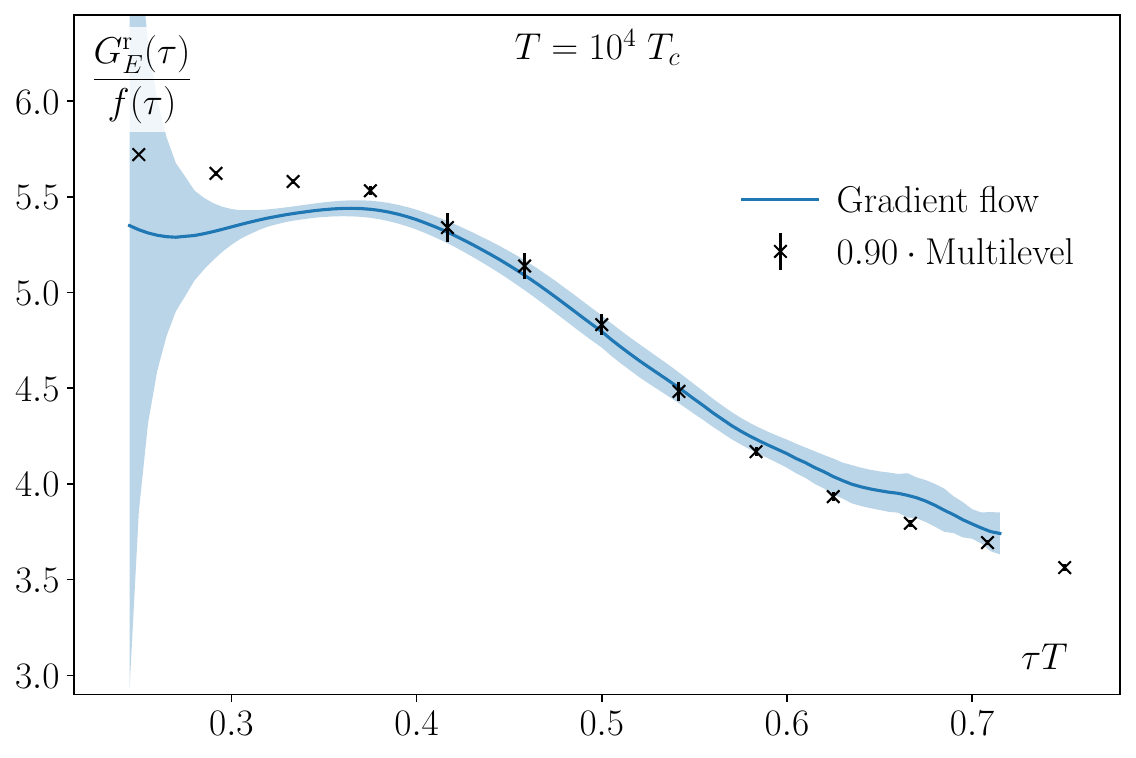}
    \caption{Final result of the non-symmetric correlator normalized with Eq.~\eqref{eq:cont_normalization}. We compare the result with results from multilevel calculations.}
    \label{fig:G_E_r_final}
\end{figure}

The final result of $G_E^r(\tau)$ is shown in Fig.~\ref{fig:G_E_r_final}. We observe that it is not symmetric around $\tau T = 0.5$, where the asymmetry is stronger at $T=1.5T_c$ than at $T=10^4T_c$. We compare the correlator from our computations using gradient flow with results from multilevel computations. For the multilevel computations, the chromoelectric fields are tadpole-improved, and the same Wilson renormalization is applied as for the gradient flow approach. We now need a multiplicative factor for the multilevel result to achieve a better agreement with the gradient flow results; the factor is closer to 1 at high temperature. Apart from the multiplicative constant, the shapes of the correlator from both approaches agree well. The lower $\tau T$-dependence, the smaller asymmetry, and the multiplicative factor closer to 1 for the normalized correlator at high temperature indicate a convergence of the non-perturbative lattice results to perturbative results. Nevertheless, the asymmetry requires a new approach to extract $\kappa^\mathrm{non-sym}$ since the established methods require a symmetric correlator.

\section{Conclusion}
We used gradient flow to measure adjoint chromoelectric correlators on the lattice at finite temperatures. We obtain two symmetric correlators, which are scaled versions of the symmetric fundamental correlator, indicating for the diffusion coefficients $\kappa^\mathrm{oct}=(5/4)\kappa^\mathrm{fund}$ and $\kappa^\mathrm{sym}=(C_A/C_F)\kappa^\mathrm{fund}$. Additionally, we obtain a non-symmetric correlator and compare it with results from multilevel computations. Nevertheless, the asymmetry requires the development of new methods to extract a reliable value for $\kappa^\mathrm{non-sym}$. The final conclusion, error analysis, and comparison with perturbation theory - see Ref.~\cite{Brambilla:2025xnw} - can be found in the full study~\cite{Brambilla:2025cqy}.

\section*{Acknowledgements}
I thank Karl Hughes for proofreading this proceedings. The simulations were carried out on the computing facilities of the Computational Center for Particle and Astrophysics (C2PAP) in the project 
\emph{Calculation of finite $T$ QCD correlators} (pr83pu) and of the SuperMUC cluster at the Leibniz-Rechenzentrum (LRZ) in the project 
\emph{Static force and other operators with field strength tensor insertions} (pn49lo), 
both located in Munich (Germany), and on the computing facilities of the Department of Theoretical Physics, TIFR.
The author gratefully acknowledge the Gauss Centre for Supercomputing e.V. 
(\href{www.gauss-centre.eu}{www.gauss-centre.eu})
for funding this project by providing computing time on the GCS Supercomputer SuperMUC-NG 
at Leibniz Supercomputing Centre (\href{www.lrz.de}{www.lrz.de}). 
This research was funded by the Deutsche Forschungsgemeinschaft (DFG, German Research Foundation) cluster of excellence "ORIGINS" (\href{https://www.origins-cluster.de}{www.origins-cluster.de}) under Germany's Excellence Strategy EXC-2094-390783311.
J.M.-S.
acknowledge the DFG Grant No. BR 4058/2-2. J.M.-S. acknowledges support by the Munich Data Science Institute
(MDSI) at the Technical University of Munich (TUM) via the Linde/MDSI Doctoral Fellowship program.

\bibliographystyle{JHEP}
\bibliography{kappa.bib}

\end{document}